# Albanian Sign Language (AlbSL) Number Recognition from Both Hand's Gestures Acquired by Kinect Sensors

Eriglen Gani

Department of Computer Science,
Faculty of Natural Sciences, University of Tirana
Tirana, Albania

Alda Kika

Department of Computer Science
Faculty of Natural Sciences, University of Tirana
Tirana, Albania

*Abstract*—Albanian Sign Language (AlbSL) is relatively new and until now there doesn't exist a system that is able to recognize Albanian signs by using natural user interfaces (NUI). The aim of this paper is to present a real-time gesture recognition system that is able to automatically recognize number signs for Albanian Sign Language, captured from signer's both hands. Kinect device is used to obtain data streams. Every pixel generated from Kinect device contains depth data information which is used to construct a depth map. Hands segmentation process is performed by applying a threshold constant to depth map. In order to differentiate signer's hands a K-means clustering algorithm is applied to partition pixels into two groups corresponding to each signer's hands. Centroid distance function is calculated in each hand after extracting hand's contour pixels. Fourier descrip-tors, derived form centroid distance is used as a hand shape representation. For each number gesture there are 15 Fourier descriptors coefficients generated which represent uniquely that gesture. Every input data is compared against training data set by calculating Euclidean distance, using Fourier coefficients. Sign with the lowest Euclidean distance is considered as a match. The system is able to recognize number signs captured from one hand or both hands. When both signer's hands are used, some of the methodology processes are executed in parallel in order to improve the overall performance. The proposed approach achieves an accuracy of 91% and is able to process 55 frames per second.

## I. INTRODUCTION

Sign Language is very important for the inclusion of the hearing impaired persons in the society. They use sign language as natural way of communication. Every country has developed their own sign language and Albania has its own, which is relatively new [1], [2]. In all situations where deaf people are participating an interpreter is required which results in a non-effective method because it requires time and resources. An Albanian Sign Language recognition system would make possible the communication between hearing impaired persons and the hearing ones in a more effective way. Many countries have tried to develope a sign language translation system through natural user interfaces as for American Sign Language [3], Arabic Sign Language [4], Portugal Sign Language [5], Indian Sign Language [6] and many others.

Until now there doesn't exist a system that is able to recognize Albanian signs by using natural user interfaces (NUI). Body movements, head position, facial expressions and hands trajectory are used by hearing impaired persons to communicate with each other. Many work has been done to integrate some existing technologies to capture and translate signer's gestures, among them web cameras [7], data gloves [8] and Kinect sensors. Web cameras generate low quality of images and have an inability to capture other body parts. It is also hard to generalize the algorithms for web cameras due to many different shapes and colors of hands. Data gloves achieve high performance but are expensive and not a proper way to human-computer interaction perspective [2]. Kinect technology, launched by Microsoft, has many advantages as: provide color and depth data simultaneously, it is inexpensive, the body skeleton can be obtained easily and it is not effected by the light. Various researchers are using Microsoft Kinect sensor for sign language recognition as in [6], [9], [10]. We are trying to built a real-time, automatic system that is able to capture and translate numbers from 1 to 10 by using Microsoft Kinect sensors. It is used to obtain input data, K-Means algorithm to differentiate signer's hands and then Fourier descriptors algorithm to classify number gestures from 1 to 10.

Many researchers have tried to capture and translate num-ber gestures by following different approaches. [11] and [12] use color camera to capture input gestures and then SVM (Support Vector Machine) and Fuzzy C-Means respectively to classify hand gestures.

[13] took another approach for number recognition. It used circle method to count the fingers which is scale, translation, and rotation invariant. Firstly the center of the hand is located and then the furthest pixel from the center is found. They both form the radius of the circle. An imaginary circle is drawn and the fingers intersecting with the circle have been counted. The method is limited to count the number of the fingers from 1 to 5 and cannot be extended to other numbers and signs.

Shape plays an important role in gesture categorization. Two approaches have been widely used for shape recognition and shape normalization including Fourier descriptors and HU moments. [14] conduced an experiment where Fourier descriptors and HU moments were compared in terms of shape recognition accuracy. The result shows that Fourier descriptors were superior to HU moments in terms of shape recognition accuracy.

Fourier descriptors can be derived from different shape signature including complex coordinates, centroid distance, curvature signature and cumulative angular function. An evaluation is done in [15], which results that centroid distance is significantly better than other three signatures. The fact that centroid distance captures both local and global features makes it desirable as shape representation.

Section I gives a brief introduction and related work. The rest of the paper is organized as follows. Section II presents an overview of methodology and a brief description of each methodology's processes. Section III describes the experimental environment. Section IV presents the experiments and results. The paper is concluded in Section V by presenting the conclusions and further work.

## II. METHODOLOGY

Figure 1 summarizes the methodology followed in number recognition captured from both signer's hands gestures:

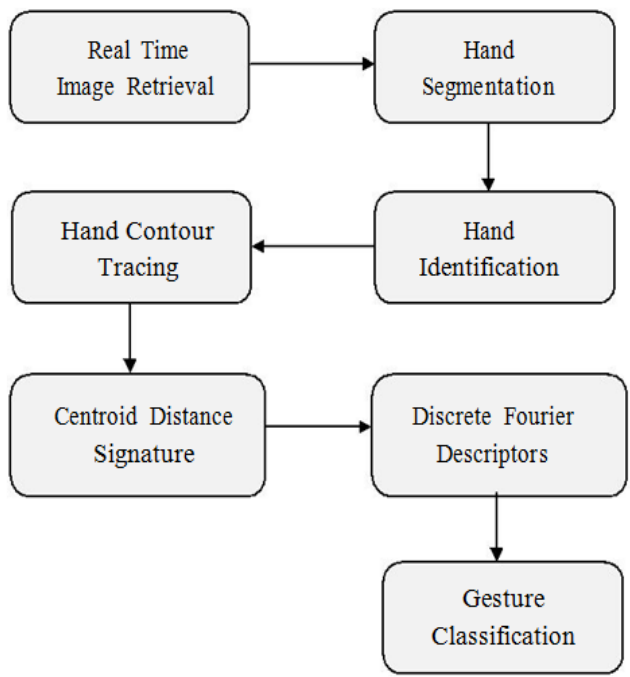

Fig. 1. Number recognition methodology

Microsoft Kinect is used to acquire hand gestures at rate of 30 frames per second. Attached to each frame it includes depth information at an interval of layers [0-4095] [16]. Each frame has dimensions of 640 x 480 (width x height) and includes a total of 307200 pixels.

Hand segmentation can be obtain by using two approaches illustrated at [17]. The first approach builds an histogram by including the number of white pixels grouped by depth layer. Histogram helps to understand the first object placed in front of Kinect. By applying a threshold constant, only the human hands are segmented. This approach does not function properly if an external object is placed between Kinect device and human body. To overcome this problem the second approach is focused only in body parts by using Kinect skeleton feature. It does not need to built an histogram because the skeleton feature excludes every object not part of human body. The first layers correspond always to user's hands. By applying again a threshold constant the human's hands are obtained.

In a 2D space the K-Means algorithm is applied in order to partition all pixels into two groups which corresponds to signer's hands. The K-means algorithm starts by placing K points (centroids) at random locations in 2D space. In our case only two centroid have been placed which correspond to user's hands. Each pixel, is assigned to a cluster with the nearest centroid, and then the new centroids are calculated as the mean of pixels assigned to it. The algorithm continues until no pixel change cluster membership. If the distance between two centroids is less then a constant, then they are merged into one.

After hand identification, the hand's contours pixels have to be extracted. A 8-connectivity algorithm [18] has been applied. The pixels that form the hand boundary are used as input data to centroid function.

Fourier descriptors are derived from shape signature. A compression between different shape signature is given at [15]. In our case centroid shape signature is used which is a one-dimensional function that represent two dimensional areas or boundaries. It is applied to pixels obtained from hand's shape boundary. Before applying Fourier transform the normalization process is performed. For matching purpose the training data set and input data must have the same number of points in the shape boundary. Number of points and points chosen affect the matching accuracy. Fast Fourier Transform (FFT) need a power-of-two [19] number of points. In our experiments 128 points have been chosen. Decreasing the number of points increases the computational results and decreases the accuracy of matching results. The chosen points must be distributed equally in all shape boundaries. There exist some methods including a)equal points sampling, b)equal angle sampling, and c)equal arclength sampling. In our experiment equal angle sampling is chosen.

Every input data must be compared against every sign in gesture dictionary by calculating Euclidean distances. The sign with the lowest Euclidean distance is considered as a match. The gesture dictionary is composed with ten number gestures representing numbers 1 to 10. Figure 2 visualizes number gestures dictionary, based on Albania Sign Language [1].

## III. EXPERIMENT ENVIRONMENT

The environment where all tests are executed is composed of a Microsoft Kinect device and a DELL Notebook. Kinect device consist of an IR emitter, an RGB camera, an IR depth sensor, a microphone array and a tilt [20]. Kinect sensors are used to obtain skeleton stream and a depth map with a resolution of 640 x 480 at 30 FPS. DELL Notebook consists of a 64-bit architecture, a Windows 7 operating system, 4 GB of physical memory and an Intel Core i5-5200U processor 2.20GHz. System is developed using Microsoft C# programming language and Kinect for Windows SDK 1.8 library. Figure 3 visualizes the system environment.

## IV. EXPERIMENT AND RESULTS

Experiments are based in two aspects: accuracy and computational latency. Firstly two data sets have been created, corresponding to training and testing data set. Each number gesture in both data sets contains 15 Fourier descriptors coefficients which are generated by running each of methodology processes.

| Number Gesture | Meaning | Number Gesture | Meaning |
|---|---|---|---|
| | One | | Six |
| | Two | | Seven |
| | Three | | Eight |
| | Four | | Nine |
| | Five | | Ten |

Fig. 2.   Number gestures dictionary

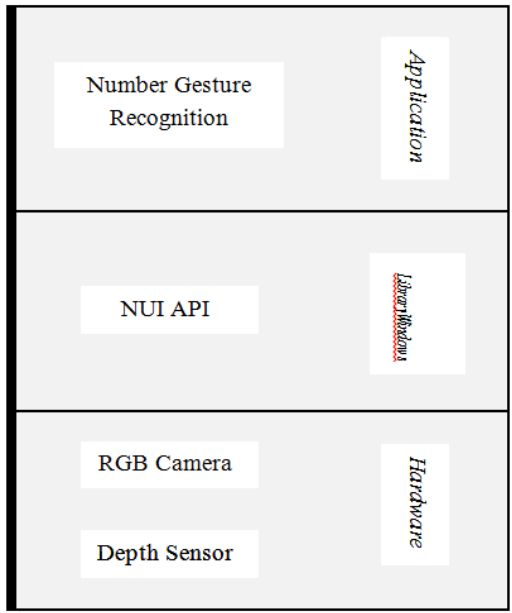

Fig. 3.   System environment

Training data set contains in total 40 gestures only for numbers 1 to 5. Numbers 6 to 10, which are generated from both user's hands are calculated as sum of numbers recognized from each individual hand. There are 8 gestures for each number. To improve accuracy, they are taken from 4 different signers using their both hands. Testing data set contains a total of 400 gestures. There are 40 gesture for each number. They are captured from 4 different signers.

TABLE I.   Testing Data Set Recognition Rate

| Number Gesture | True Recognition | False Recognition |
|---|---|---|
| 1 | 100% | 0% |
| 2 | 88% | 12% |
| 3 | 95% | 5% |
| 4 | 85% | 15% |
| 5 | 93% | 7% |
| 6 | 95% | 5% |
| 7 | 93% | 7% |
| 8 | 85% | 15% |
| 9 | 93% | 7% |
| 10 | 85% | 15% |
| Average | 91% | 9% |

Figure 4 shows the confusion matrix for number recognition system. Gestures 10, 8 and 4 have the lowest recognition rate. They are misclassfied as gestures 9, 7 and 5. Number 9 is the most ambiguous gesture even through having a high recognition rate of 93%. It is recognized as gestures 7, 8 and 10.

| | 1 | 2 | 3 | 4 | 5 | 6 | 7 | 8 | 9 | 10 |
|---|---|---|---|---|---|---|---|---|---|---|
| 1 | 40 | 0 | 0 | 0 | 0 | 0 | 0 | 0 | 0 | 0 |
| 2 | 3 | 35 | 2 | 0 | 0 | 0 | 0 | 0 | 0 | 0 |
| 3 | 0 | 0 | 38 | 0 | 0 | 2 | 0 | 0 | 0 | 0 |
| 4 | 0 | 0 | 1 | 34 | 4 | 0 | 0 | 0 | 0 | 0 |
| 5 | 0 | 0 | 0 | 2 | 37 | 1 | 0 | 0 | 0 | 0 |
| 6 | 0 | 0 | 0 | 0 | 1 | 38 | 0 | 0 | 0 | 1 |
| 7 | 0 | 0 | 0 | 0 | 0 | 2 | 37 | 0 | 0 | 1 |
| 8 | 0 | 0 | 0 | 0 | 0 | 0 | 6 | 34 | 0 | 0 |
| 9 | 0 | 0 | 0 | 0 | 0 | 0 | 1 | 1 | 37 | 1 |
| 10 | 0 | 0 | 0 | 0 | 0 | 0 | 0 | 1 | 5 | 34 |

Fig. 4.   Confusion matrix

Firstly training data set elements are used as input data. The system receives 100% recognition rate. Secondly testing data set elements are used as input. Table I summarizes the results. Numbers 10, 8 and 4 have the lowest recognition rate while numbers 1, 3, and 6 have the highest recognition rate.

Computational latency is firstly calculated for each process running sequentially. Hand segmentation and K-Means calculation are the heaviest processes which take approximately 60% of total time. If all the processes are executed in sequentially order the system is able to calculate 48 frames per second. Microsoft Kinect is able to recognize number gestures in real-time since it generates frames at a rate of 30 frames per second. Table II summarizes the results. Time in milliseconds for each process is the average time of 50 experiments where single and both hands have been used.

Today's processors are implemented with a dozen of cores and this number is going to increase due to Moore's law [21]. Multi-core processors allow parallelism and multithread-ing. Since performance is an important factor in real-time recognition systems, computation latency is recalculated in situation where some level of parallelism is provided to some of processes. It is clear that some of the processes can be processed in parallel. K-Means Calculation, Hand Contour Tracing, Normalize Image (128 points), Centroid Distance Sig-nature, Discrete Fourier Description and Gesture Classification can all be processed in parallel for each signer's hand. Then the results must be aggregated. Table III summarizes the results. Time in milliseconds for each process is the average time of 50 experiments where single and both hands have been used. The total time is improved by 2.9 milliseconds, and the system is able to process 55 frames per second.

TABLE II.    COMPUTATION LATENCY CALCULATED SEQUENTIALLY

| Processes | Time in milliseconds |
|---|---|
| Hand Segmentation | 6.4381 |
| K-Means Calculation | 6.0661 |
| Hand Contour Tracing | 4.2646 |
| Normalize Image (128 points) | 0.0574 |
| Centroid Distance Signature | 0.0671 |
| Discrete Fourier Description | 3.7005 |
| Gesture Classification | 0.3477 |
| Total | 20.9415 |

TABLE III.    COMPUTATION LATENCY CALCULATED FROM PARALLEL TASKS

| Processes | Time in milliseconds |
|---|---|
| Hand Segmentation | 6.4579 |
| K-Means Calculation | 5.6461 |
| Hand Contour Tracing | 3.2571 |
| Normalize Image (128 points) | 0.0434 |
| Centroid Distance Signature | 0.0503 |
| Discrete Fourier Description | 2.5015 |
| Gesture Classification | 0.1397 |
| Total | 18.0960 |

## V.    CONCLUSION AND FUTURE WORK

Aim of this paper is to propose an automatic, real-time solution for recognition of a limited set of numbers (1 to 10) obtained from signer's both hands, by using Microsoft Kinect. This system can be extended in the future to include more signs (including dactyls and not static signs) creating the first sign recognition system for Albanian Sign Language by using natural user interfaces (NUI). Input images are provided through Microsoft Kinect device. IR depth sensor is used to built a depth map. By using depth map the hand segmentation is performed. Skeleton feature of Kinect device can be used in hand segmentation process. In order to understand if gesture is formed from single hand or both hands a K-Means algorithm is applied. The system is able to recognize gestures capture from single hand or both hands by switching automatically. For each segmented hand the contour pixels are extracted. Each number gesture is represented by 15 Fourier descriptors coefficients which are based on centroid distance signature. In total, data set consists of 440 number gestures where 40 of them are used to form training data set. 400 number gestures are used to form testing data set. Every gesture in testing data set is compared against each gesture in training data set by using Euclidean distance. The gesture with minimum distance is considered as a match. Numbers 10, 8, and 4 have the lowest recognition rate. They are misclassified as gestures 9, 7 and 5. Number 9 is the most ambitious one. The system achieves an accuracy of 91%. Computation latency allows the system to be deployed in an image receiving technology that has an acquisition rate of less than 48 frames per second where no parallelism is applied and 55 frames per second where parallelism is applied. Microsoft Kinect can be part of this real-time recognition system since it's acquisition rate is 30 frames per second.

Future work consists of improving the overall system accu-racy by applying more reliable gesture data set and improving the execution time of the slowest processes. Future work will be extended to dactyls and other not static sign gestures.

REFERENCES

[1]    ANAD, Gjuha e Shenjave Shqipe 1, ANAD, Ed. Shoqata Kombetare Shiptare e Njerezve qe nuk Degjojne, 2013.

[2]    E. Gani and A. Kika, "Review on natural interfaces technologies for designing albanian sign language recognition system," The Third Inter-national Conference On: Research and Education Challenges Towards the Future, 2015.

[3]    F. Ullah, "American sign language recognition system for hearing impaired people using cartesian genetic programming," in Automation, Robotics and Applications (ICARA), 2011 5th International Conference on. IEEE, 2011, pp. 96–99.

[4]    N. R. Albelwi and Y. M. Alginahi, "Real-time arabic sign language (arsl) recognition," in International Conference on Communications and Information Technology (ICCIT 2012), Tunisia, 2012, pp. 497–501.

[5]    P. Trindade and J. Lobo, "Distributed accelerometers for gesture recog-nition and visualization," in Technological Innovation for Sustainability. Springer, 2011, pp. 215–223.

[6]    A. S. Ghotkar and G. K. Kharate, "Dynamic hand gesture recognition and novel sentence interpretation algorithm for indian sign language using microsoft kinect sensor," Journal of Pattern Recognition Research, vol. 1, pp. 24–38, 2015.

[7]    S. Shruthi, K. Sona, and S. Kiran Kumar, "Classification on hand gesture recognition and translation from real time video using svm-knn," International Journal of Applied Engineering Research, vol. 11, no. 8, pp. 5414–5418, 2016.

[8]    R. Rupasinghe, D. Ailapperuma, P. De Silva, A. Siriwardana, and B. Sudantha, "A portable tool for deaf and hearing impaired people."

[9]    K. Stefanov and J. Beskow, "A kinect corpus of swedish sign language signs," in Proceedings of the 2013 Workshop on Multimodal Corpora: Beyond Audio and Video, 2013.

[10]    H. V. Verma, E. Aggarwal, and S. Chandra, "Gesture recognition using kinect for sign language translation," in Image Information Processing (ICIIP), 2013 IEEE Second International Conference on. IEEE, 2013, pp. 96–100.

[11]    J. Wachs, U. Kartoun, H. Stern, and Y. Edan, "Real-time hand gesture telerobotic system using fuzzy c-means clustering," in Automation Congress, 2002 Proceedings of the 5th Biannual World, vol. 13. IEEE, 2002, pp. 403–409.

[12]    Y. Liu, Z. Gan, and Y. Sun, "Static hand gesture recognition and its application based on support vector machines," in Software Engineering, Artificial Intelligence, Networking, and Parallel/Distributed Computing, 2008. SNPD'08. Ninth ACIS International Conference on. IEEE, 2008, pp. 517–521.

[13] A. Malima, E. Ozgur,˜ and M. C¸etin, "A fast algorithm for vision-based hand gesture recognition for robot control," in Signal Processing and Communications Applications, 2006 IEEE 14th. IEEE, 2006, pp. 1–4.

[14] S. Conseil, S. Bourennane, and L. Martin, "Comparison of fourier descriptors and hu moments for hand posture recognition," in Signal Processing Conference, 2007 15th European. IEEE, 2007, pp. 1960–1964.

[15] D. Zhang, G. Lu et al., "A comparative study of fourier descriptors for shape representation and retrieval," in Proc. 5th Asian Conference on Computer Vision. Citeseer, 2002.

[16] J. Webb and J. Ashley, Beginning Kinect Programming with the Microsoft Kinect SDK. Apress, 2012.

[17] E. Gani and A. Kika, "Identifikimi i dores nepermjet teknologjise microsoft kinect," Buletini i Shkencave te Natyres, vol. 20, pp. 82–90, 2015.

[18] T. Pavlidis, Algorithms for graphics and image processing. Springer Science & Business Media, 2012.

[19] J. W. Cooley and J. W. Tukey, "An algorithm for the machine calculation of complex fourier series," Mathematics of computation, vol. 19, no. 90, pp. 297–301, 1965.

[20] MSDN, "Kinect for windows sensor components and specifcations," April 2016. [Online]. Available: https://msdn.microsoft.com/en-us/library/jj131033.aspx

[21] G. E. Moore, "Cramming more components onto integrated circuits," Readings in computer architecture, vol. 56, pp. 56–59, 2000.